\documentclass[12pt,a4paper]{article}
\usepackage[applemac]{inputenc}
\usepackage{amssymb}
\usepackage{amsmath}
\usepackage{amsopn}
\newcommand{\LL}{\mathbb{L}}
\newcommand{\CC}{\mathbb{C}}

\newcommand{\BB}{\mathbb{B}}

\newcommand{\KK}{\mathbb{K}}

\newcommand{\NN}{\mathbb{N}}

\newcommand{\RR}{\mathbb{R}}

\newcommand{\frE}{\mathfrak{E}}
\newcommand{\frF}{\mathfrak{F}}

\newcommand{\frZ}{\mathfrak{Z}}

\newcommand{\frm}{\mathfrak{m}}

\newcommand{\kaa}{\mathcal{A}}

\newcommand{\kD}{\mathcal{D}}
\newcommand{\kE}{\mathcal{E}}

\newcommand{\kh}{\mathcal{H}}

\newcommand{\kj}{\mathcal{J}}
\newcommand{\kK}{\mathcal{K}}
\newcommand{\kL}{\mathcal{L}}
\newcommand{\mm}{\mathcal{M}}

\newcommand{\kP}{\mathcal{P}}

\newcommand{\kR}{\mathcal{R}}

\newcommand{\gk}{\kappa}
\newcommand{\gl}{\lambda}

\newcommand{\tm}{\subseteq}

\newcommand{\∞}{\infty}
\newcommand{\ten}{\otimes}
\newcommand{\Ds}{\bigoplus}
\newtheorem{definition}{Definition}[section]
\newtheorem{proposition}{Proposition}[section]
\newtheorem{theorem}{Theorem}[section]
\newtheorem{lemma}{Lemma}[section]
\newtheorem{corollary}{Corollary}[section]
\newtheorem{remark}{Remark}[section]
\newtheorem{example}{Example}[section]

\newcommand{\por}{\kP_{0}(\kR)}

\newcommand{\pr}{\kP(\kR)}

\newcommand{\lh}{\mathcal{L}(\mathcal{H})}

\newcommand{\all}{\forall}

\newcommand{\rr}{\kR}
\newcommand{\er}{\kE(\kR)}
\newcommand{\hr}{\kR_{sa}}

\newcommand{\eal}{E^{A}_{\gl}}

\newcommand{\ebl}{E^{B}_{\gl}}
\newcommand{\epl}{E^{P}_{\gl}}
\newcommand{\eql}{E^{Q}_{\gl}}
\newcommand{\eakl}{E^{A_{\gk}}_{\gl}}
\newcommand{\ea}{E^{A}}
\newcommand{\eak}{E^{A_{\gk}}}
\newcommand{\eb}{E^{B}}

\newcommand{\we}{\wedge}
\newcommand{\We}{\bigwedge}
\newcommand{\Ve}{\bigvee}
\newcommand{\eao}{E^{A}_{0}}
\newcommand{\ebo}{E^{B}_{0}}
\newcommand{\tto}{\mapsto}

\newcommand{\smm}{\setminus}

\newcommand{\dr}{\kD(\rr)}

\newcommand{\dprr}{\kD_{pr}(\rr)}

\newcommand{\irr}{\in \RR}
\newcommand{\lir}{\gl \in \RR}

\newcommand{\kik}{k \in \KK}

\begin{document}
\title{\Huge{On a canonical lattice structure \\ on the effect algebra of a von 
Neumann algebra}}

\author{Hans F.\ de Groote\footnote{e-mail: degroote@math.uni-frankfurt.de}
	 \\ FB Mathematik\\ J.W.Goethe-Universit\"{a}t\\
	 Frankfurt a.\ M.}

\date{14.12.2005}
\maketitle
\bibliographystyle{plain}

\begin{abstract}
    Let $\kR$ be a von Neumann algebra acting on a Hilbert space 
    $\kh$  and let $\kR_{sa}$ be the set of hermitean (i.e. 
    selfadjoint) elements of $\kR$. It is well known that $\kR_{sa}$ 
    is a lattice with respect to the usual partial order $≤$ if and only 
    if $\kR$ is \emph{abelian}. We define and study a new partial 
    order on $\kR_{sa}$, the spectral order $≤_{s}$, which extends
    $≤$ on projections, is coarser than the usual one, but agrees with
    it on abelian subalgebras, and turns $\kR_{sa}$ into a
    \emph{boundedly complete lattice}. The effect algebra $\er := \{ A
    \in \hr | 0 ≤ A ≤ I \}$ is then a complete lattice and we show
    that the mapping $A \mapsto R(A)$, where $R(A)$ denotes the range 
    projection of $A$, is a homomorphism from the lattice $\er$ onto
    the lattice $\pr$ of projections if and only if $\rr$ is a finite 
    von Neumann algebra.    
    
 \end{abstract}
 \pagebreak
 
 \section{Introduction}
 \label{sec: A}
 
 In this paper $\kR$ is a von Neumann algebra contained in the 
 algebra $\lh$ of bounded linear operators of the Hilbert space 
 $\kh$, $\kR_{sa}$ denotes the set of hermitean elements of $\kR$ and 
 $\er$ the \emph{effect algebra of $\kR$}, i.e.  set of all positive
 operators in $\kR$ less or equal to $I$. This is meant with respect to the usual
 partial order on $\kR_{sa}$:
 \[
     A ≤ B \quad \text{if and only if} \quad \forall \ x \in \kh : \ <Ax, 
     x> \ ≤ \  <Bx, x>. 
 \]
 
 \begin{definition}\label{A}
 A \emph{lattice} is a partially ordered set $(\LL, \leq)$  such 
that any two elements $a,b \in\LL$ possess a \emph{maximum} $a\vee b 
\in\LL$ and a \emph{minimum} $a\wedge b \in\LL$.\\
Let $\frm$ be an infinite cardinal number.\\
The lattice $\LL$ is called $\mathfrak{m}$-complete, if every family 
$(a_{i})_{i\in I}$ has a supremum $\bigvee_{i \in I}a_{i}$ and an 
infimum $\bigwedge_{i\in I}a_{i}$ in $\LL$, provided that $\# I\leq 
\mathfrak{m}$ holds.
A lattice $\LL$ is simply called complete, if every family 
$(a_{i})_{i \in I}$ in $\LL$ (without any restriction of the 
cardinality of $I$) has a maximum and a minimum in $\LL$.\\
$\LL$ is said to be boundedly complete if every bounded family in 
$\LL$ has a maximum and a minimum. \\
If a lattice has a \emph{zero element} $0$ ( i.e. $\forall a \in \LL : 
0\leq a$) and a \emph{unit element} $1$ (i.e. $\forall a\in \LL : a\leq 
1$) then completeness and bounded completeness are the same.\\
Note that a complete lattice always has a zero and a unit element,
namely $0 := \We_{a \in \LL}a$ and $1 := \Ve_{a \in \LL}a$. \\
A lattice $\LL$ is called \emph{distributive} if the two distributive 
laws
\begin{eqnarray*}
	a \wedge (b \vee c) & = & (a \wedge b) \vee (a \wedge c)  \\
	a \vee (b \wedge c) & = & (a \vee b) \wedge (a \vee c)
\end{eqnarray*}
hold for all elements $a, b, c \in \LL$. 
\end{definition}
$\bigvee_{i \in I}a_{i}$ is characterized by the following universal 
property:
\begin{enumerate}
	\item [(i)]  $\forall j\in I :\quad a_{j}\leq \bigvee_{i \in I}a_{i}$

	\item [(ii)]  $\forall c\in \LL :\quad ((\forall i\in I : a_{i} \leq c) 
	\Rightarrow \bigvee_{i}a_{i} \leq c ).$
\end{enumerate}
An analogous universal property characterizes the minimum 
$\bigwedge_{i}a_{i}$.\\
~\\
Note that if  $\LL$ is a distributive complete lattice, then in 
general
\begin{displaymath}
	a \wedge (\bigvee_{i \in I}b_{i}) \ne \bigvee_{i \in I}(a \wedge 
	b_{i}),
\end{displaymath}
so completeness and distributivity do not imply \emph{complete 
distributivity}!\\
~\\
Now a well known theorem states (\cite {[KadRing3]},p.186):
\begin{theorem}\label{thm: B}
    Let $\kaa$ be a $C^{*}-$algebra. Then $\kaa_{sa}$ is a lattice with 
    respect to the partial order $≤$ if and only if $\kaa$ is abelian.
\end{theorem}
But note:
\begin{itemize}
    \item [(i)] The relation $≤$ is closely connected with the linear 
    structure of $\kaa_{sa}$: $A ≤ B$ if and only if $B -A ≥ 0$, 
    whereas 

    \item  [(ii)] $A ≥ 0$ can also be characterized by the fact that 
    the spectrum $sp(A)$ of $A$ is contained in $\RR_{+}$, the set of 
    nonnegative real numbers.
\end{itemize}
In this paper we will, based on the foregoing observation, define a new
partial order $≤_{s}$, called the {\bf spectral order} on $\kR_{sa}$, 
and study its main properties. \\
~\\
After publishing the first version of this paper in the arXiv, David
Sherman (\cite{sh}) informed me that the definition of the spectral
order and its main properties are already contained in a paper of M.P.
Olson (\cite{ol}). Because I came to these results in a more general
context (\cite{dg}), and because the first version contains an
important application of the spectral order, I think that it is
justified to publish this second version. \\
~\\
The spectral order can be defined by elementary relations between the
spectral projections of the operators in question. The spectral order
agrees (by its very definition) on projections with the usual one but,
in general, it is coarser than that. This means that $A ≤_{s} B$ implies
$A ≤ B$ but not vice versa. It turns out that the two partial orders
agree for all commuting pairs of hermitean operators $A, B$, a fact
that should be important for possible applications in quantum physics.
In section \ref{sec: C} we show that the spectral order turns $\hr$ into
a boundedly complete lattice $(\hr \vee_{s}, \we_{s})$. This is
equivalent to the completeness of the sublattice $\er$. There is a
natural hull operation on the effect algebra $\er$:
\begin{eqnarray*}
    R: \er & \to & \pr  \\
    A & \mapsto & R(A),
\end{eqnarray*}
where $R(A)$ denotes the range projection of $A$. $R$ always respects 
the join $\vee_{s}$. It respects also the meet $\we_{s}$ if and only
if $\rr$ is a \emph{finite} von Neumann algebra. A similar result has 
been obtained by C. Cattaneo and J. Hamhalter in \cite{[CatHam]} -
though for the usual order, where $\vee$ and $\we$ are only partially
defined operations.      
\pagebreak

\section{The spectral order}% contains main definition and comparison 
%with usual partial order
\label{B}

Let $\pr$ be the lattice of projections in the von Neumann algebra 
$\kR$ and let $\por := \pr \smm \{0\}$. We have introduced in
\cite{dg} the notion of \textit{observable function} of an element 
$A \in \hr$. This is a bounded real valued function $f_{A}$ on the
space $\dr$ of all \emph{dual ideals} of the lattice $\pr$, defined by
\[
    \all \ \kj \in \dr : \ f_{A}(\kj) := \inf \{ \gl \irr \mid \eal
    \in \kj \},
\]
where $(\eal)_{\lir}$ is the spectral family of $A$. The
restriction of $f_{A}$ to the set $\dprr$ of all \textit{principal
dual ideals} $H_{P} := \{ Q \in \pr \mid Q ≥ P \}, \ (P \in \por)$,
defines a function
\[
    \begin{array}{cccc}
        r_{A} : & \por & \to & \RR  \\
         & P & \tto & f_{A}(H_{P}).
     \end{array}
\]  
The functions $r : \por \to \RR$ that are induced by observable
functions are characterized by the property that
\[
    r(\Ve_{\kik}P_{k}) = \sup_{\kik}r(P_{k})
\]
holds for all families $(P_{k})_{\kik}$ in $\por$. Therefore, they are
called \emph{completely increasing functions} (\cite{dg}). \\
~\\
The following result is easy to prove:

\begin{proposition}\label{b1}
    Let $A, B \in \hr$ with spectral families $\ea$ and $\eb$,
    respectively. Then 
    \[
        r_{A} ≤ r_{B} \quad \text{if and only if} \quad \all \ \gl \irr :  \ \ebl ≤
        \eal.
    \]
\end{proposition}
~\\  
One can reconstruct $f_{A}$ from $r_{A}$, because  
\[
    \all \ \kj \in \dr : \ f_{A}(\kj) = \inf_{P \in \kj}r_{A}(P).
\]
The spectral family of $P \in \pr$ is given by
\[
    E^{P}_{\gl} = 
    \begin{cases}
        0        & \text{for $\gl < 0$}\\
        I - P    & \text{for $0 ≤ \gl < 1$}\\
        I         & \text{for $1≤ \gl$}.
    \end{cases}
\]
If $P, Q \in \pr$, then $P ≤ Q$ if and only if $I - Q ≤ I - P$ i.e.
\[
    P ≤ Q \quad \text{if and only if} \quad \forall \ \gl \in \RR: \ 
    E^{Q}_{\gl} ≤ E^{P}_{\gl} . 
\]  
These simple facts lead us to the following basic

\begin{definition}\label{def:Ba}
    Let $A, B \in \kR_{sa}$ with corresponding spectral families 
    $\ea = (\eal)_{\gl \in \RR}$ and $\eb = (\ebl)_{\gl \in \RR}$, 
    respectively. Then $A ≤_{s} B$ if and only if
    \[
        \forall \ \gl \in \RR: \ \ebl ≤ \eal.  
    \]  
    $≤_{s}$ is a partial order on $\hr$. It is called the {\bf
    spectral order}.
\end{definition}              

\begin{remark}\label{b2}
    The mapping $A \tto r_{A}$ from $\hr$ onto the set of completely
    increasing functions is not additive. Therefore we can not expect 
    that $≤_{s}$ is a linear order. Hence the spectral order should be
    different from the usual one.
\end{remark}
In the sequel we will investigate the relations between the spectral 
order and the usual order on $\kR_{sa}$. To this end we show that we 
can confine ourselves to the subset $\er$ of hermitean operators 
between $0$ and $I$. This makes the discussion somewhat more 
comfortable.

\begin{lemma}\label{lem: Bb}
    Let $a, b \in \RR, a > 0$. Then for all $A, B \in \kR_{sa}$ the 
    following equivalences hold:
    \begin{enumerate}
        \item  $A ≤ B \quad \Longleftrightarrow \quad aA + bI ≤ aB + bI$,
        
        \item  $A ≤_{s} B \quad \Longleftrightarrow \quad aA + bI 
        ≤_{s} aB + bI.$
    \end{enumerate}
\end{lemma}
\emph{Proof}: The first equivalence is trivial. The second follows 
from the simple fact that the spectral family $E^{aA + bI}$ of $aA + bI$ 
is given by
\[
    E^{aA + bI}_{\gl} = E^{A}_{a^{-1}\gl - b},
\]
where $E^{A}$ is the spectral family of $A$:
\[
    E^{aB + bI}_{\gl} = E^{B}_{a^{-1}\gl - b} ≤ E^{A}_{a^{–1}\gl - b} = 
    E^{aA + bI}_{\gl}.
\]
$\Box$\\
~\\
The following example (which is taken from \cite{[KadRing1]}, p.251)
shows that the two partial orders on $\kR_{sa}$ are different.
\begin{remark}\label{rem: Bc}
    Let $\kh = \CC^2$ and $P = 
    \begin{pmatrix}
        1 & 0 \\
        0 & 0
    \end{pmatrix},
    A = 
    \begin{pmatrix}
        2 & 1 \\
        1 & 1
    \end{pmatrix}.$
    Then $P ≤ A$, but $P \nleq_{s} A$.
\end{remark}
 \emph{Proof:} A simple calculation shows $P ≤ A$. $A$ has 
 eigenvalues $\gl_{1} = \frac{3}{2} - \frac{1}{2} \sqrt{5} < 1$ and
 $\gl_{2} = \frac{3}{2} + \frac{1}{2} \sqrt{5} > 1$. Therefore 
 $E^{A}_{\gl_{1}}$ 
 is the projection  onto the line $\CC \begin{pmatrix}
 \frac{1}{2}(1- \sqrt{5}) \\ 1
 \end{pmatrix}$,
 but $E^{P}_{\gl_{1}}$ is the projection onto the line
 $\CC \begin{pmatrix} 0 \\ 1
 \end{pmatrix}.$
Hence \  $E^{A}_{\gl_{1}} \nleq  E^{P}_{\gl_{1}}$, i.e. $P \nleq_{s} A$.
$\Box$\\

\begin{theorem} \label{theo: Bd}
    The spectral order on $\kR_{sa}$ is coarser than the usual one, i.e.
    \[
        \all \ A, B \in \kR_{sa}: \ (A ≤_{s} B \ \Longrightarrow \ 
        A ≤ B).
    \]
\end{theorem}
\emph{Proof:} According to \ref{lem: Bb} we can assume that $A, B \in 
\er$. By the spectral theorem $A$  and $B$ are ( in norm) 
arbitrarily close to
\[
    \sum_{k = 1}^{n}\frac{k}{n}(E^{A}_{\frac{k}{n}} - 
    E^{A}_{\frac{k - 1}{n}})
\] 
 and 
 \[
     \sum_{k = 1}^{n}\frac{k}{n}(E^{B}_{\frac{k}{n}} - 
     E^{B}_{\frac{k - 1}{n}})
 \]
 respectively if $n$ is chosen sufficiently large. Because of 
 $-E^{A}_{\gl} ≤ -E^{B}_{\gl}$ for all $\gl$ we obtain  
 \begin{eqnarray*}
     \sum_{k = 1}^{n}\frac{k}{n}(E^{A}_{\frac{k}{n}} - 
        E^{A}_{\frac{k - 1}{n}}) & = & I - \frac{1}{n}(E^{A}_{\frac{n - 1}{n}} + 
        E^{A}_{\frac{n - 2}{n}} + \ldots + E^{A}_{0})      \\
      & ≤ &  I - \frac{1}{n}(E^{B}_{\frac{n - 1}{n}} + 
        E^{B}_{\frac{n - 2}{n}} + \ldots + E^{B}_{0})     \\
      & = & \sum_{k = 1}^{n}\frac{k}{n}(E^{B}_{\frac{k}{n}} - 
        E^{B}_{\frac{k - 1}{n}}). 
 \end{eqnarray*}
 Hence $A ≤ B.$
 $\Box$\\

 \begin{corollary}\label{cor: Be}
    If $A, B \in \kR_{sa}$ commute, then
    \[
        A ≤_{s} B \ \Longleftrightarrow \ A ≤ B.
    \] 
 \end{corollary}
 \emph{Proof:} The spectral projection $E^{A}_{\gl}$ is the projection 
 onto the kernel of $(A - \gl I)^{+}$. Therefore, if $A ≤ B$ and if 
 $A$ and $B$ commute, it follows that
 \[
 \all \ \gl: \ (A - \gl I)^{+} ≤ (B - \gl I)^{+}.
 \]
 Hence $ker (B - \gl I)^{+} \tm ker (A - \gl I)^{+}$, i.e. 
 $E^{B}_{\gl} ≤ E^{A}_{\gl}$ for all $\gl$.
 $\Box$\\
 
 \begin{remark}\label{rem: Bea}
     If $A, B \in \kR_{sa}$, then $A ≤_{s} B$ does not imply that $A$ 
     and $B$ commute: Let $B \in \er$ be invertible, $P$ an 
     arbitrary projection. Then $aP ≤_{s} B$ for sufficiently small 
     $a > 0$, but $aPB \neq BaP$ in general. 
 \end{remark}
 
 \begin{remark}\label{rem: Beb}
     The proof of the foregoing corollary shows the core of the
     difference between the two partial orders: if $A$ and $B$ are two
     {\bf noncommuting} hermitean operators, then $A ≤ B$ does {\bf
     not} imply the relation $A^+ ≤ B^+$ for their positive parts. The
     reason for this is that the function $t \mapsto t^2$ is not
     operator-monotonic. Indeed T. Ogasawara has shown (\cite{[Og]}, see 
     also \cite{[Ham]}, theorem 7.3.4) that a $C^{*}$-algebra $\kaa$ with
     the property
     \[
         \all \ a, b \in \kaa : \ ( 0 ≤ a ≤ b \quad \Longrightarrow
	\quad a^2 ≤ b^2 )
     \]
     is necessarily abelian.
 \end{remark}
 
 \begin{corollary}\label{cor: Bf}
     Let $A, B \in \er$ such that $A$ or $B$ is a projection. 
     Then
     \[
         A ≤_{s} B \ \Longleftrightarrow \ A ≤ B.
     \] 
 \end{corollary}
 \emph{Proof:} According to corollary \ref{cor: Be} we only have to show that 
 $A ≤ B$ forces $A$ and $B$ to commute. We may assume that $A$ is a 
 projection $P$, because $I - B ≤ I - A$ reduces the other possibility 
 to the first one.\\
 Now $P ≤ B$ implies
 \[
     P ≤ PBP ≤ PIP = P,
 \]
i.e.
\[
    P = PBP.
\]
Therefore $B$ leaves $im P$ invariant: Let $x$ be a unit vector from
$im P$. Then we can write $Bx = y + z$ with $y \in im P, \ z \in (im
P)^\perp$. Because of $|y|^2 + |z|^2 = |Bx|^2 ≤ 1$ and
\[
    x = Px = PBx = y,
\]
$z = 0$ follows. As $B$ is hermitean, $(im P)^{\perp}$ is
$B$-invariant, too. This shows $PB = BP$, and from $P = PBP$ we even 
get $P = PB = BP$.
$\Box$\\
~\\
\emph{Note that the example in Remark \ref{rem: Bc} shows that the assumption 
$A, B \in \er$ is essential in the foregoing corollary.}\\
~\\
We close this section with a short comment on a possible physical
interpretation of the spectral order. \\
Let $\ea$ be the spectral family of $A \in \er$. Then
\[
    I - \eal = \chi_{]\gl, 1]}(A)
\]
where $\chi_{]\gl, 1]}$ denotes the characteristic function of the
interval $]\gl, 1]$. Hence 
\[
    \all \ A, B \in \er : \ ( A  ≤_{s} B \quad \Longleftrightarrow
    \quad  \all \ \gl \in [0, 1] : \ \chi_{]\gl, 1]}(A) ≤ \chi_{]\gl, 1]}(B). 
\]   
If $x$ is a unit vector in $\kh$ then $<\chi_{]\gl, 1]}(A)x, x>$ is
usually interpreted as the probability that measuring the observable
$A$ in the pure state $x$ gives a result lying in the interval $]\gl, 
1]$.

\section{The spectral lattice}
\label{sec: C}

% Enthält die Definition der Verbandsoperationen und den Nachweis 
% ihrer beschränkten Vollständigkeit.

In this section we show that $\hr$ is a boundedly complete lattice
with respect to the spectral order. In order to motivate our definitions we 
reconsider the lattice operations for projections.\\
If $P, Q \in \pr$ then 
\[
    I - (P \vee Q) = (I - P) \we (I - Q) \quad \text{and} \quad I - 
    (P \we Q) = (I - P) \vee (I - Q).
\]
Therefore the spectral families of $P \vee Q$ and $P \we Q$ are given 
by
\[
      E^{P \vee Q}_{\gl} = \epl \we \eql
\]
and 
\[
      E^{P \we Q}_{\gl} = \epl \vee \eql
\]
respectively. This leads to the following generalization.

\begin{proposition}\label{prop: Ca}    
    Let $\frE = (E_{\gl})_{\gl \in \RR}$ and $\frF = (F_{\gl})_{\gl \in 
    \RR}$ be spectral families in $\rr$. Then
    \begin{itemize}
        \item [(i)] $(\frE \vee \frF)_{\gl} := E_{\gl} \we F_{\gl} 
        \qquad (\gl \in \RR)$  and\\
        
        \item [(ii)] $(\frE \we \frF)_{\gl} := \We_{\mu > 
        \gl}(E_{\mu} \vee F_{\mu}) \qquad (\gl \in \RR)$ 
     \end{itemize}
    define spectral families $\frE \vee \frF$ and $\frE \we \frF$ 
    respectively in $\rr$.
\end{proposition}
\emph{Proof:} The only not totally trivial point is the continuity of 
$\frE \we \frF$ from the right:
\begin{eqnarray*}
    \We_{\nu > \gl}(\frE \we \frF)_{\nu} & = & \We_{\nu > \gl} 
    \We_{\mu > \nu}(E_{\mu} \vee F_{\mu})  \\
     & = & \We_{\mu > \gl}(E_{\mu} \vee F_{\mu})  \\
     & = & (\frE \we \frF)_{\gl}. \qquad \Box
\end{eqnarray*}
At a first glance the infimum over $\mu$ in the definition of 1$\frE 
\we \frF$ looks strange but it is necessary in order to guarantee 
the continuity from the right. This is shown by the following

\begin{example}\label{ex: Cb}
    Let $\kh$ be separable, $(e_{k})_{k \in \NN}$ an orthonormal 
    basis for $\kh$, $x := \sum_{k = 1}^{\∞}\frac{1}{k}e_{k}$, $P$ the 
    projection onto $\CC x$ and $P_{n}$  the projection onto
    \[
        U_{n} = \CC e_{1} + \ldots + \CC e_{n}.
    \]
    Note that $x \notin U_{n}$ for all $n \in \NN$. Let $\frE = 
    (E_{\gl})_{\gl \in \RR}$ be the spectral family defined by
    \[
        E_{\gl} :=
        \begin{cases}
	0      & \text{for $\gl ≤ 0$}\\
	I - P_{n}  & \text{for $\frac{1}{n + 1} ≤ \gl < \frac{1}{n}$}\\
	I       & \text{for $\gl ≥ 1$}
        \end{cases}
    \]
    and let $\frF$ be the spectral family of $P$. Then 
    \[
        (\frE \we \frF)_{0} = \We_{\mu > 0}(E_{\mu} \vee (I - P)).
    \]
    As $x \notin U_{n}$ we obtain for $\mu \in [\frac{1}{n+1}, 
    \frac{1}{n}[$:
    \begin{eqnarray*}
        E_{\mu} \vee (I - P) & = & (I - P_{n}) \vee (I - P)  \\
         & = & I - (P_{n} \we P)  \\
         & = & I.
    \end{eqnarray*}
    Hence $(\frE \we \frF)_{0} = I$, but $E_{0} \vee (I - P) = I - 
    P < I$.
\end{example}

\begin{remark}\label{rem: Cba}
    If $\rr$ is a finite von Neumann algebra and if
    $(P_{\iota})_{\iota \in J}, (Q_{\iota})_{\iota \in J}$ are
    decreasing nets (over the same index set $J$) in $\pr$ converging 
    to projections $P$ and $Q$ respectively, then (see
    \cite{[KadRing4]}, p.412)
    \[
        \We_{\iota \in J}(P_{\iota} \vee Q_{\iota}) = P \vee Q.
    \]
    Thus for finite $\rr$ we have
    \[
        (\frE \we \frF)_{\gl} = E_{\gl} \vee F_{\gl} \quad \text{for
        all} \quad \gl \in \RR.
    \]
\end{remark}

\begin{definition}\label{def: Cc}
    Let $A, B \in \hr$ with corresponding spectral families $\ea$ 
    and $\eb$ respectively. Then we define $A \we_{s} B, \ A \vee_{s} B$ as 
    the operators in $\hr$ whose spectral families are $\ea \we \eb$ 
    and $\ea \vee \eb$ respectively. 
\end{definition}

\begin{proposition}\label{prop: Cd}
    $A \we_{s} B$ is the minimum and $A \vee_{s} B$ is the maximum of $A$ 
    and $B$ in the sense of lattice theory.
\end{proposition}
\emph{Proof:} We have to check the universal properties of minimum and 
maximum. \\
$A \we_{s} B ≤_{s} A$, for
\[
    \all \ \gl \in \RR: \ E^{A \we_{s} B}_{\gl} = \We_{\mu > 
    \gl}(E^{A}_{\mu} \vee E^{B}_{\mu}) ≥ \We_{\mu > 
    \gl}E^{A}_{\mu} = \eal 
\]
and similarly $A \we_{s} B ≤_{s} B$. If $C \in \hr$ such that $C ≤_{s} 
A, B$, then
\[
    E^{A}_{\mu}, E^{B}_{\mu} ≤ E^{C}_{\mu},
\]
hence
\[
    E^{A}_{\mu} \vee E^{B}_{\mu} ≤ E^{C}_{\mu}
\]
for all $\mu \in \RR$, and therefore 
\[
    \We_{\mu > \gl}(E^{A}_{\mu} \vee E^{B}_{\mu}) ≤ \We_{\mu > 
    \gl}E^{C}_{\mu} = E^{C}_{\gl}.
\]   
This shows $C ≤_{s} A \we_{s} B$. In the same way one can prove that 
$A, B ≤_{s} A \vee_{s} B$ and that $A, B ≤_{s} C$ implies $A \vee_{s} 
B ≤_{s} C$. $\Box$\\  
~\\
Thus $\hr$ together with the spectral order $≤_{s}$ is a lattice 
which we call the {\bf spectral lattice of $\rr$} and denote it by 
$(\hr, ≤_{s})$. If we speak of the \emph{lattice} $\hr$, we always 
mean this with respect to the spectral order. (There cannot be any 
confusion with the usual order: if $\rr$ is not abelian, then $\hr$ 
is not a lattice with respect to $≤$, and if $\rr$ is abelian, then 
the two partial orders coincide.) From corollary \ref{cor: Bf} we 
obtain 
\begin{corollary}\label{cor: Ce}
    The projection lattice $\pr$ is a sublattice of the spectral 
    lattice $\hr$.
\end{corollary}

\begin{lemma}\label{lem: Cf}
    For $A \in \hr$ let $[m(A), M(A)]$ be the smallest compact
    interval containing the spectrum $sp(A)$ of$A$. Then for all $A, B
    \in \hr$
    \begin{eqnarray*}
        m(A \we_{s} B) & = & min (m(A), m(B)),  \\
        M(A \we_{s} B) & ≤ & min (M(A), M(B)),  \\
        m(A \vee_{s} B)  & ≥ & max (m(A), m(B)),  \\
        M(A \vee_{s} B) & = & max (M(A), M(B)).
    \end{eqnarray*}
\end{lemma}
This is quite easy to see and so we omit the proof.\\
~\\
From this lemma and from lemma \ref{lem: Bb} we further obtain 

\begin{corollary}\label{cor: Cg}
    For $a, b \in \RR, \ a < b,$ let
    \[
        \rr_{[a, b]} := \{A \in \hr \mid aI ≤ A ≤ bI \}.   
    \]
    Then $(\rr_{[a, b]}, ≤_{s})$ is a sublattice of the spectral
    lattice $\hr$, isomorphic to $\er$.
\end{corollary}

\begin{theorem}\label{theo: Ch}
    The spectral lattice $\hr$ is boundedly complete.
\end{theorem}
\emph{Proof:} Obviously, $\hr$ is a boundedly complete lattice if and 
only if $\er$ is a complete lattice. We prove the completeness of
$\er$. Let $(A_{\gk})_{\gk} \in \KK$ be an arbitrary family in $\er$
and let $(\eak)_{\gk \in \KK}$ be the corresponding family of spectral
families. Let
\[
    \all \ \gl \in \RR : \ E^{\vee}_{\gl} := \We_{\gk \in \KK}\eakl.
\]
It is quite easy to check that $\frE^{\vee} := (E^{\vee}_{\gl})_{\gl \in \RR}$ is a 
spectral family and that the corresponding operator $A_{\frE^{\vee}}$
belongs to $\er$. From the definition of $\frE^{\vee}$ we have 
$A_{\gk} ≤_{s}A_{\frE^{\vee}}$ 
for all $\gk \in \KK$. Let $B \in \er$ with spectral family $\frF =
(F_{\gl})_{\gl \in \RR}$ such that $A_{\gk} ≤_{s} B$ for all $\gk \in 
\KK$, i.e.
\[
    \all \ \gk \in \KK \ \all \ \gl \in \RR : \ F_{\gl} ≤ \eakl. 
\] 
Hence
\[
    \all \ \gl \in \RR : \ F_{\gl} ≤ \We_{\gk \in \KK}\eakl, 
\]
i.e. $A_{\frE^{\vee}} ≤_{s} B.$ Therefore
\[
    \Ve_{\gk \in \KK}A_{\gk} := A_{\frE^{\vee}}
\]
is the \emph{supremum of the family $(A_{\gk})_{\gk \in \KK}$.}\\
In order to show that $(A_{\gk})_{\gk \in \KK}$ has an infimum we set
\[
     E^{\we}_{\gl} := \We_{\mu > \gl}\Ve_{\gk \in \KK}E^{A_{\gk}}_{\mu}.
\]
We show that $\frE^{\we} := (E^{\we}_{\gl})_{\gl \in \RR}$ is a
spectral family.\\
The properties $E^{\we}_{\gl} = 0$ for $\gl < 0$ and $E^{\we}_{\gl} = 
1$ for $\gl ≥ 1$ are obvious.\\
Let $\gl_{1} < \gl_{2}$ and $\mu, \nu$ such that $\gl_{1} < \mu <
\gl_{2} < \nu.$ Then $\Ve_{\gk}E^{A_{\gk}}_{\mu} ≤
\Ve_{\gk}E^{A_{\gk}}_{\nu}$ and therefore
\[
     \Ve_{\gk}E^{A_{\gk}}_{\mu} ≤ \We_{\nu >
     \gl_{2}}\Ve_{\gk}E^{A_{\gk}}_{\nu}.
\] 
This implies
\[
    \We_{\mu > \gl_{1}}\Ve_{\gk}E^{A_{\gk}}_{\mu} ≤ \We_{\nu >
    \gl_{2}}\Ve_{\gk}E^{A_{\gk}}_{\nu},
\]
i.e.
\[
    E^{\we}_{\gl_{1}} ≤ E^{\we}_{\gl_{2}}.
\]
Finally we have
\begin{eqnarray*}
    \We_{\mu > \gl}E^{\we}_{\mu} & = & \We_{\mu > \gl}\We_{\nu >
    \mu}\Ve_{\gk}E^{A_{\gk}}_{\nu}  \\
     & = & \We_{\mu > \gl}\Ve_{\gk}E^{A_{\gk}}_{\mu}  \\
     & = & E^{\we}_{\gl}.
\end{eqnarray*}
Hence $\frE^{\we}$ is a spectral family in $\er$. Eventually we prove 
that the operator $A_{\frE^{\we}}$ corresponding to this spectral
family is the infimum of $(A_{\gk})_{\gk \in \KK}$.\\
Let $B \in \er$ with spectral family $\frF = (F_{\gl})_{\gl} \in \RR$ 
such that $B ≤ A_{\gk}$ for all $\gk \in \KK$. Then
$\Ve_{\gk}E^{A_{\gk}}_{\mu} ≤ F_{\mu}$ for all $\mu$, hence
\[
    \all \ \nu > \gl : \ \We_{\mu > \gl}\Ve_{\gk}E^{A_{\gk}}_{\mu} ≤
    F_{\nu}
\] 
and therefore
\[
    E^{\we}_{\gl} = \We_{\mu > \gl}\Ve_{\gk}E^{A_{\gk}}_{\mu}  
     ≤ \We_{\nu > \gl}F_{\nu}  = F_{\gl}
\]     
for all $\gl$. Thus
\[
    B ≤ A_{\frE^{\we}}. \qquad \Box
\]

\section{Complements}
\label{sec: D}

We have defined the spectral order and the corresponding lattice
operations in terms of spectral families. In the same way we proceed
to define complementations.\\
If $\frE = (E_{\gl})_{\gl \in \RR}$ is a spectral family in the von
Neumann algebra $\rr$ then $\gl \mapsto I - E_{- \gl}$ is 
increasing but it is not necessarily continuous from the right:
\begin{eqnarray*}
    \We_{\mu > \gl}(I - E_{- \mu}) & = & \We_{\mu < - \gl}(I -
    E_{\mu})  \\
     & = & I - \Ve_{\mu < - \gl}E_{\mu}  \\
     & = & I - E_{- \gl -0},
\end{eqnarray*}    
where $E_{a - 0} := \Ve_{\mu < a}E_{\mu}.$ 
This leads to the definition
\[
    (\neg \frE)_{\gl} := \We_{\mu > \gl}(I - E_{- \mu}) = I - E_{- \gl
    - 0}.
\]
It is easy to check that $\neg \frE$ is a spectral family. We call it 
the {\bf free complement of $\frE$}.

\begin{proposition}\label{prop: Da}
    Let $A \in \hr$ with spectral family $\ea$. Then $\neg \ea = E^{- 
    A}$.
\end{proposition}
\emph{Proof:} The hermitean operator given by $\neg \ea$ is
\[
    \neg A := \int_{\RR}\gl d(\neg \ea)_{\gl}.
\]   
From $(\neg \ea)_{\gl} = I - E^{A}_{- \gl - 0}$ we obtain
\[
    \neg A = \int_{\RR}\gl d(- \ea_{- \gl - 0}). 
\] 
If $\frZ = (\gl_{k})_{k \in K}$ is a partition of $\RR$, then
\[
    \sum_{k}\mu_{k}(- \ea_{- \gl_{k + 1} - 0} + \ea_{- \gl_{k} - 0}) = -
    \sum_{k}(- \mu_{k})(\ea_{- \gl_{k} - 0} - \ea_{- \gl_{k + 1} - 0})
\]
where $\mu_{k} \in ]\gl_{k}, \gl_{k + 1}[$. This converges to $- A$ as
the width $|\frZ|$ of $\frZ$ tends to zero because of
\[
    \ea([a, b[) = \ea_{b - 0} - \ea_{a - 0} \quad \text{and} \quad
    \ea(]a, b]) = \ea_{b} - \ea_{a}.
\]
Hence $\neg A = - A$.  \ \ $\Box$

\begin{corollary}\label{cor: Db}
    $\neg(\neg \frE) = \frE$ for all spectral families $\frE$ in $\rr$.
\end{corollary}
If $A \in \hr$ then obviously
\[
    [m(- A), M(- A)] = [- M(A), - m(A)],
\]
so

\begin{remark}\label{rem: Dc}
    If $A \in \rr_{[a, b]}$ then $(a + b)I - A \in \rr_{[a, b]}$.
    Especially $I - A \in \er$ for $A \in \er$.
\end{remark}
If $A \in \er$ then $I - A$ is called the {\bf Kleene complement}
(\cite{[CatHam]}). If $A$ is a projection, then $A \we (I - A) = 0$.
This is not true for general $A \in \er$:

\begin{proposition}\label{prop: Dd}
    Let $A \in \er$. Then $A \we (I - A) = 0$ if and only if $A$ is a 
    projection.
\end{proposition}
This is a well known result with a quite simple proof: Consider $A, I 
- A$ as continuous functions $sp(A) \to [0, 1]$. If $A(\gl) > 0$ then
$A \we (I - A) = 0$ implies $1 - A(\gl) = 0$, i.e. $A(\gl) = 1$. This 
means that $im A \tm \{0, 1\}$, i.e. that $A$ is a projection.\\
~\\
The Kleene complement satisfies the de Morgan rules in the lattice
$\er$:

\begin{proposition}\label{prop: De}
    Let $A, B \in \er$. Then
    \begin{enumerate}
        \item [(i)] $A ≤_{s} B$ if and only if $I - B ≤_{s} I - A$,  
    
        \item [(ii)] $I - (A \we_{s} B) = (I - A) \vee_{s} (I - B)$, 
    
        \item [(iii)] $I - (A \vee_{s} B) = (I - A) \we_{s} (I - B)$. 
    \end{enumerate}
\end{proposition}
\emph{Proof:} $A ≤_{s} B$ implies
\[
    E^{I - A}_{\gl} = I - E^{A}_{(1 - \gl) - 0} ≤ I - E^{B}_{(1 - \gl)
    - 0} = E^{I - B}_{\gl} 
\]
for all $\gl$, hence $I - B ≤_{s} I - A$.\\
From the universal property of the maximum we conclude that
$I - (A \we_{s} B) = (I - A) \vee_{s} (I - B)$ if and only if
\[
    \all \ C \in \er : \ (I - A ≤_{s} C, I - B ≤_{s} C \Longrightarrow
    I - (A \we_{s} B) ≤_{s} C).
\]
This follows from (i) and the universal property of the minimum: 
\begin{eqnarray*}
    I - A ≤_{s} C, I - B ≤_{s} C & \Longrightarrow & I - C ≤_{s} A, I 
    - C ≤_{s} B  \\
     & \Longrightarrow & I - C ≤_{s} A \we_{s} B  \\
     & \Longrightarrow & I - (A \we_{s} B) ≤_{s} C.
\end{eqnarray*}
(iii) follows from (i) and (ii). \ \ $\Box$ \\

\begin{corollary}\label{cor: Dea}
    Let $A \in \er$. Then $A \vee (I - A) = I$ if and only if $A$ is a
    projection.
\end{corollary}
If $A \in \rr$ then the projection onto  the closure of $im A$ is
called the {\bf range projection of $A$} and is usually denoted by
$R(A)$. Obviously
\[
    R(A) = \We \{ P \in \pr | \ PA = A \}.
\] 

\begin{lemma}\label{lem: Df}
    Let $A \in \er$. Then 
    \[
        R(A) = \We \{ P \in \pr | \ A ≤ P \} = I - \eao. 
    \]
\end{lemma}
\emph{Proof:} If $A \in \er$ and $P \in \pr$, then $PA = A$ implies
\[
    A = PA = PAP ≤ P.
\] 
Conversely $A ≤ P$ implies $A = PA$ by the proof of corollary \ref{cor: 
Bf}. Hence $R(A) = \We \{ P \in \pr | \ A ≤ P \}$. \\
If $P$ is a projection in $\rr$ then $A ≤ P$ is equivalent to $A ≤_{s}
P$. This is equivalent to $I - P ≤ \eao$ i.e. to $I - \eao ≤ P$.
Therefore $R(A) = I - \eao$. \ \ $\Box$  \\

\begin{definition}\label{def: Dg}(\cite{[CatHam]})
    $A^{\sim} := I - R(A) = \eao$ is called the {\bf Brouwer complement of
    $A \in \er$}. 
\end{definition}
The Brouwer complement has the following (certainly well known)
properties:

\begin{proposition}\label{prop: Dh}
    For all $A \in \er$ the following properties hold:
    \begin{enumerate}
        \item [(i)] $A^{\sim\sim} = R(A)$, 
    
        \item [(ii)] $A \we A^{\sim} = 0$, but 
    
        \item [(iii)] $A \vee A^{\sim} = I$ if and only if $A$ is a
        projection. 
    \end{enumerate}
\end{proposition}
\emph{Proof:} $(i)$ is obvious. $A \we (I - R(A)) ≤ R(A) \we (I -
R(A)) = 0$ gives $(ii)$. $(iii)$ follows from
\[
    E^{A \vee A^{\sim}}_{\gl} = \eal \we (I - \eao) = \eal - \eao 
\]
for all $\gl \in [0, 1[$. \ \  $\Box$ \\
~\\
Eventually we will show that the {\bf hull operation} $R : \er \to 
\pr$ which sends $A$ to its range projection $R(A)$ is a lattice
homomorphism if and only if $\rr$ is a finite von Neumann algebra.

\begin{lemma}\label{lem: Di}
    For all $A, B \in \er$ we have
    \begin{enumerate}
        \item  [(i)] $R(A \vee_{s} B) = R(A) \vee R(B)$,  
    
        \item  [(ii)] $R(A \we_{s} B) ≤ R(A) \we R(B)$.
    \end{enumerate}
\end{lemma}
\emph{Proof:} Although these properties follow immediately from
lemma \ref{lem: Df}  and the universal property of the lattice
operations, it is instructive to give a proof that uses the definition
of the lattice operations.\\
$E^{A \vee_{s} B}_{0} = \eao \we \ebo$ implies 
\[
    R(A \vee_{s} B) = I - \eao \we \ebo = (I - \eao) \vee (I - \ebo) =
    R(A) \vee R(B),
\]
and
\[
    E^{A \we_{s} B}_{0} = \We_{\gl > 0}(\eal \vee \ebl) ≥ \eao \vee
    \ebo
\]
implies
\[
    R(A \we_{s} B) = I - E^{A \we_{s} B}_{0} ≤ I - (\eao \vee \ebo) = 
    (I - \eao) \we (I - \ebo) = R(A) \we R(B). \ \ \Box   
\]

\begin{remark}\label{rem: Dj}
    The proof gives the essential hint for proving that $R(A \we_{s}
    B) = R(A) \we R(B)$ for all $A, B \in \er$ forces the finiteness
    of $\rr$. Example \ref{ex: Cb} shows that in $\lh$ the strict
    inequality $R(A \we_{s} B) < R(A) \we R(B)$ can occur and we shall
    construct a similar example in any non-finite von Neumann algebra.
\end{remark}

\begin{corollary}\label{cor: Dk}
    The Brouwer complement satisfies the first de Morgan law
    \[
        \all \ A, B \in \er : \ (A \vee_{s} B)^{\sim} = A^{\sim} \we
        B^{\sim}.
    \]
\end{corollary}

\begin{lemma}\label{lem: Dl}
    Let $\mm \tm \kL(\kK)$ be a von Neumann algebra acting on a
    Hilbert space $\kK$ with unity $I_{\mm} = id_{\kK}$ and let $(E_{\gl})_{\gl
    \in \RR}$ be a spectral family in the von Neumann algebra $\rr \tm
    \lh$. Then for all $A, B \in \hr$ and all $P, Q \in \pr$:
    \begin{enumerate}
        \item  [(i)] $I_{\mm} \otimes A ≤ I_{\mm} \otimes B$ if and
        only if $A ≤ B$. 
    
        \item  [(ii)] $I_{\mm} \ten (P \we Q) = (I_{\mm} \ten P) \we
        (I_{\mm} \ten Q)$.
    
        \item  [(iii)] $I_{\mm} \ten (P \vee Q) = (I_{\mm} \ten P) \vee
        (I_{\mm} \ten Q)$.
    
        \item  [(iv)] $(I_{\mm} \ten E_{\gl})_{\gl \in \RR}$ is a
        spectral family in $\mm \bar{\ten} \rr$.     
    \end{enumerate}
\end{lemma}
\emph{Proof:} We use some results on tensor products that can be found
in \cite{[KadRing1], [KadRing2], [Tak1]}.\\
Let $(e_{b})_{b \in \BB}$ be an orthonormal basis of $\kK$. Then 
\[
    U : \sum_{b \in \BB}x_{b} \mapsto \sum_{b \in \BB}(e_{b} \ten
    x_{b})
\]
is a surjective isometry from $\Ds_{b \in \BB}\kh_{b}$ (with
$\kh_{b} = \kh$ for all $b \in \BB$) onto $\kK \ten \kh$. Let $A \in
\rr$. $U$ intertwines $I_{\mm} \ten A$ and $A$ :
\[
    U^{-1}(I_{\mm} \ten A)U = \Ds_{b \in \BB}A_{b}
\] 
with $A_{b} = A$ for all $b \in \BB$. This immediately implies
$(i)$.\\
Note that $I_{\mm } \ten A$ is a projection if and only if $A$ is.
Then $(ii)$ and $(iii)$ follow from $(i)$ and the universal property
of minimum and maximum.\\
Let $(E_{\gl})_{\gl \in \RR}$ be a spectral family in $\rr$. Then $\gl
\mapsto I_{\mm} \ten E_{\gl}$ is monotonic increasing, equals $I_{\mm}
\ten I$ for $\gl$ large enough and zero for $\gl$ small enough. In
order to prove the continuity from the right we use the fact that the 
mapping $A \mapsto I_{\mm} \ten A$ from $\rr$ to $\mm \bar{\ten} \rr$ 
is strongly continuous on bounded subsets of $\rr$:
\[
    \We_{\mu > \gl}(I_{\mm} \ten E_{\mu}) = I_{\mm} \ten \We_{\mu >
    \gl}E_{\mu} = I_{\mm} \ten E_{\gl}.
\]
Hence also $(iv)$ follows. \ \ $\Box$

\begin{theorem}\label{theo: Dm}
    The mapping $R : \er \to \pr, \ \ A \mapsto R(A)$ is a
    homomorphism of lattices if and only if $\rr$ is a finite von
    Neumann algebra.
\end{theorem}
\emph{Proof:} Remark \ref{rem: Cba} shows that $R : \er \to \pr $ is a
lattice homomorphism if $\rr$ is finite. Now assume that $\rr$ is not 
finite. Then $\rr$ contains a direct summand of the form $\mm
\bar{\ten} \kL(\kh_{0})$, where $\mm \tm \kL(\kK)$ is a suitable von
Neumann algebra and $\kh_{0}$ a separable Hilbert space of infinite
dimension (see e.g. \cite{[Tak1]}, Ch. V.1, essentially prop. 1.22: if
$\rr$ is not finite then $\rr$ has a direct summand with properly
infinite unity $I_{0}$. Use the halving lemma to construct a countable
infinite orthogonal sequence of pairwise equivalent projections with
sum $I_{0}$ (see the proof of theorem 6.3.4 in \cite{[KadRing2]})). Take
the spectral families $(E_{\gl})_{\gl \in \RR}$ and $E^{P}$
in $\kL(\kh_{0})$ we have defined in example \ref{ex: Cb}.
Then by lemma \ref{lem: Dl}
\begin{eqnarray*}
    \We_{\mu > 0}((I_{\mm} \ten E_{\mu}) \vee (I_{\mm} \ten (I - P))) & 
    = & \We_{\mu > 0}(I_{\mm} \ten (E_{\mu} \vee (I - P)))  \\
     & = & I_{\mm} \ten \We_{\mu > 0}(E_{\mu} \vee (I - P))  \\
     & > & I_{\mm} \ten (E_{0} \vee (I - P)).
\end{eqnarray*}
Therefore we obtain for the corresponding operators $I_{\mm} \ten A$
and $I_{\mm} \ten P$:
\[
    R((I_{\mm} \ten A) \we_{s} (I_{\mm} \ten P)) < R(I_{\mm} \ten A) \we
    R(I_{\mm} \ten P),
\] 
i.e. $R$ is not a lattice homomorphism. \ \ $\Box$

\begin{corollary}\label{cor: Dn}
    The von Neumann algebra $\rr$ is finite if and only if the second 
    de Morgan law
    \[
        (A \we_{s} B)^{\sim} = A^{\sim} \vee B^{\sim} 
    \]
    for the Brouwer complement is satisfied for all $A, B \in \er$.
\end{corollary}

\vspace{1cm}

{\bf Acknowledgements.} Sincere thanks go to Andreas Döring for
several discussions on the subject and for bringing the paper of 
C.Cattaneo and J.Hamhalter to my attention. Thanks go also to David
Sherman (Santa Barbara) for informing me about Olson's paper.

\pagebreak

\end{document}